\documentclass[floats,floatfix,showpacs,amssymb,prl,twocolumn,superscriptaddress,nofootinbib]{revtex4-2}

\usepackage{graphicx}% Include figure files
\usepackage{dcolumn}% Align table columns on decimal point
\usepackage{bm}% bold math
\usepackage{amsmath}
\usepackage[colorlinks=true,linkcolor=blue,citecolor=blue,urlcolor=blue]{hyperref}
\usepackage{dsfont}

\usepackage{xcolor}
\usepackage{ulem}

\newcommand{\tildecalE}{{\tilde {\cal E}}}
\def\vev#1{{\left\langle #1 \right \rangle}}

\newcommand\comment[1]{}

\begin{document}

%\preprint{APS/123-QED}

\title{Gravothermal collapse of self-interacting dark-matter halos with anisotropic velocity distributions}

\author{Marc Kamionkowski}
\email{kamion@jhu.edu}
\affiliation{William H.\ Miller III Department of Physics \& Astronomy, Johns Hopkins University, 3400 N.\ Charles St., Baltimore, MD 21218, USA}

\author{Kris Sigurdson}
\email{krs@phas.ubc.ca}
\affiliation{Department of Physics and Astronomy, University of British Columbia, Vancouver, BC V6T 1Z1, Canada}

\setcounter{footnote}{0}
\def\thefootnote{\arabic{footnote}}

\begin{abstract}
Self-gravitating galactic halos composed of self-interacting dark matter exhibit the formation of a highly dense core at the galactic center---a gravothermal collapse.  Analytic models to describe this evolution have been developed and calibrated to numerical simulations initialized with isotropic particle velocity distributions, an assumption not necessarily warranted by the theory of halo formation.  Here we study the dependence of the timescale for gravothermal collapse on the velocity distribution.  To do so, we consider self-consistent initial conditions for halos with the same density distribution but with different velocity distributions.  We consider models with constant anisotropy and with an anisotropy that increases with radius. The velocity distributions that we explore have collapse times that differ from that assuming isotropic distributions by more than a factor of two.  We argue that these variations may depend on the global changes in velocity-dispersion profiles in these anisotropic models and not just on the degree of anisotropy.
\end{abstract}

\maketitle
%\tableofcontents

%\section{Introduction}

Self-interacting dark matter (SIDM) \cite{Tulin:2017ara,Buckley:2017ijx,Spergel:1999mh} has  been suggested as an explanation for discrepancies between observed properties of dwarf galaxies \cite{Kaplinghat:2015aga,Kamada:2016euw,Correa:2020qam,Zeng:2021ldo,Zentner:2022xux,Correa:2022dey,Yang:2022mxl,Nadler:2023nrd} and those expected if the dark-matter halo is made of collisionless matter.  SIDM provides a mechanism for heat transport in the halo.  In halos with cusps---density profiles that scale with radius $r$ as $r^{-1}$ as $r\to 0$---the heat transport initially smooths the density profile to a nearly constant value in the interior \cite{Colin:2002nk}.  With time, though, heat flow out of the center drives the system towards gravothermal core collapse, a runaway process during which the central density grows rapidly, arising from negative specific heat where the core contracts and heats while losing energy~\cite{Kochanek:2000pi,Balberg:2001qg}.  This process has been described by analytic models \cite{Koda:2011yb,Tulin:2017ara,Essig:2018pzq,Nishikawa:2019lsc,Slone:2021nqd,Outmezguine:2022bhq,Yang:2022zkd,Gad-Nasr:2023gvf,Dave:2000ar} with parameters calibrated to N-body simulations \cite{Kummer:2019yrb,Fischer:2020uxh,Mace:2024uze,Palubski:2024ibb,Fischer:2024eaz,Mace:2025fuz,Fischer:2025rky}.

The simulations are initialized using Eddington inversion formula \cite{Eddington:1916,Binney:2008}, Eq.~(\ref{eqn:eddington}) below, to assign a velocity distribution to the dark-matter particles.  This then generates a self-gravitating dark-matter halo that is stationary in time if the particles are collisionless.  In this way, the effects of the evolution of the halo with self-interactions can be understood.  The Eddington inversion formula is derived under the assumption that the distribution function (DF) is independent of angular momentum, or equivalently, that the dark-matter velocity distribution is everywhere isotropic.  However, realistic halos are not necessarily expected to have isotropic velocity distributions.  The process of spherical infall tends to favor radial orbits, a notion supported by evidence from cosmological simulations \cite{Wojtak:2013eia,He:2024gvw,Lemze_2012} and observations \cite{Biviano:2013eia}.  It is thus worth understanding the dependence of the physics of gravothermal collapse on the assumed velocity distribution.

\begin{figure*}[htbp]
\includegraphics[width=2\columnwidth]{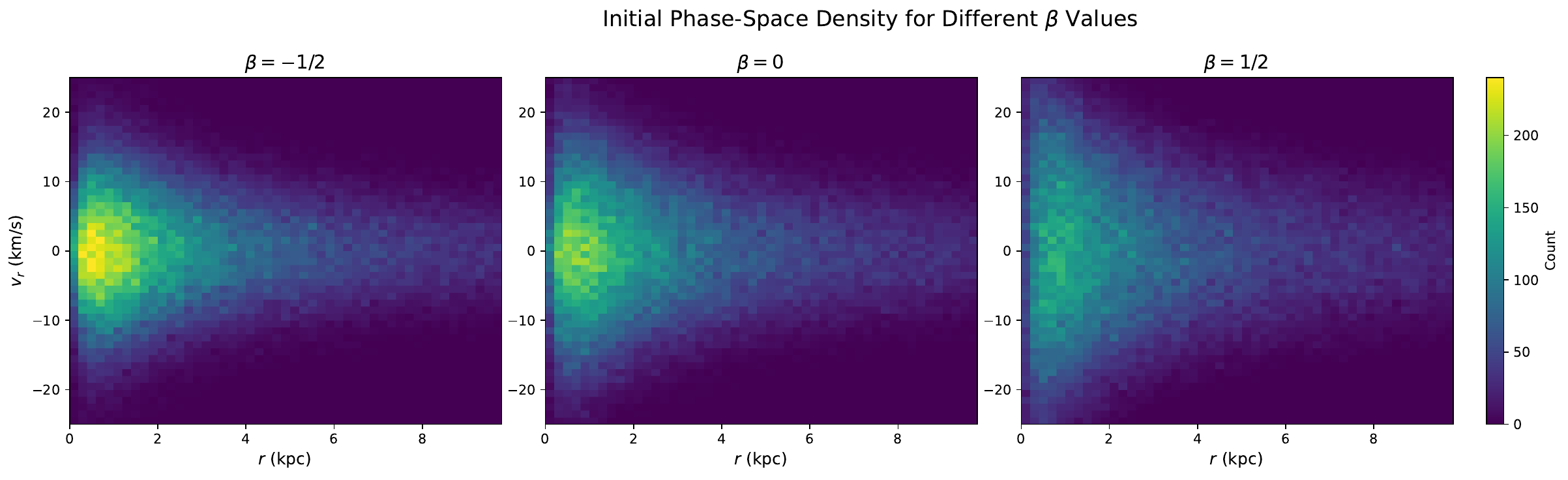}
     \caption{The initial distributions of particles in the radius and radial velocity.  The radial velocities become increasingly spread out as we increase from $\beta=-1/2$ (more tangential orbits) to $\beta=1/2$ (more radial orbits).}
\label{fig:phasespace}
\end{figure*}

Here, we use a recently developed approach to simulating self-gravitating dark-matter halos \cite{Kamionkowski:2025uae,Kamionkowski:2025fat} (see also Ref.~\cite{Gurian:2025zpc}) to study the dependence of the timescale for gravothermal collapse of SIDM halos on the assumed velocity distribution.  This approach builds upon the realization that particles in a spherical system occupy a three-dimensional subspace of the full six-dimensional phase space.  It also capitalizes upon the fact that the force on any given particle is in the radial direction and determined just by the mass at smaller radii.  These simulations are, for spherical systems, far more computationally efficient than traditional N-body simulations, and several of the pitfalls associated with traditional N-body simulations are circumvented.  

We run simulations that are initialized two different ways: some with anisotropy in the velocity distribution constant with radius and some isotropic as $r \to 0$ with anisotropy that increases with radius.  As will be seen below, the gravothermal-collapse timescales in these models can differ by more than a factor of 2.  For constant-anisotropy models, distribution functions with predominantly radial orbits lead to longer collapse times while those with tangential orbits have shorter collapse times over the range of parameters explored.  Models with initial anisotropy that increases with radius can be parametrized by the anisotropy at the scale radius of the halo. The collapse timescale does not vary monotonically with this characteristic anisotropy.  In both sets of models, the dependence of three-dimensional velocity dispersion on radius changes as the anisotropy is changed, and our results suggest that the collapse timescale is affected by the shape of this altered velocity-dispersion profile as well as the anisotropy.

We proceed by reviewing that the initial conditions for prior simulations have been obtained by drawing the positions $\bf x_i$ and velocities $\bf v_i$ (with $i=1,\ldots,N$) of the $N$ particles (of mass $m_p$) from a DF,
\begin{equation}
     f({\bf x},{\bf v}) = f_{\rm init}({\cal E}) = \frac{1}{m_p\sqrt{8}\pi^2} \frac{d}{d{\cal E}}\int_0^{\cal E}\, \frac{d\Psi}{\sqrt{{\cal E}-\Psi}} \frac{d\rho}{d\Psi},
\label{eqn:eddington}     
\end{equation}
where $\Psi(r)$ is (minus) the gravitational potential, obtained from $\nabla^2\Psi = - 4 \pi G \rho$, and ${\cal E} = \Psi-v^2/2$.  This Eddington formula guarantees that the initial density profile will be preserved in the absence of interactions \cite{Binney:2008}, but it does so by assuming that the DF is independent of the angular momentum $L$ of the particle, or equivalently, that the velocity distribution is everywhere isotropic.

However, it is possible to generate self-consistent {\it anisotropic} velocity distributions --- in fact an infinitude of such models.  Anisotropic DFs have velocity dispersions in the radial and tangential directions related by $(1-\beta)\vev{v_r^2} = \vev{v_\theta^2} = \vev{v_\phi^2}$.  Velocity distributions have a preponderance of orbits that are isotropic when $\beta=0$, radial where $\beta>0$, and tangential where $\beta<0$.  Here, in general, $\beta = \beta(r)$ where $\beta(r)$ is an arbitrary function, but when referencing specific $\beta(r)$ models hereafter it is implied that $\beta$ denotes the appropriate form for each model by context.

Here we consider a set of models with a constant anisotropy and also a set of Osipkov-Merritt (OM) \cite{Jiang:2007vsd,Ossipkov1979,Merritt1985,Dejonghe1987,Baes:2002tw,AnEvans,Binney:2008} models.  These models have an anisotropy that increases with radius.  Although realistic halos are more likely to have an anisotropy that increases with radius, we caution against thinking of the OM models as more correct as the anisotropy increases more rapidly and the orbits are far more radial at large radii than we expect in halos from cosmological simulations. The two classes of models instead bracket expected anisotropy profiles.

The constant-anisotropy models have a DF of the form $f({\cal E},L) = L^{-2\beta} f_1({\cal E})$, where $L$ is the angular momentum per unit mass.    For $-1/2<\beta<1/2$ we have \cite{Binney:2008},
\begin{equation}
    f_1({\cal E}) \propto % = \frac{C_\beta}{m_p} 
    \frac{d}{d\Psi} \int_0^\Psi \frac{d\Psi}{({\cal E}-\Psi)^{1/2-\beta}} \frac{d}{d\Psi}\left( \left[r(\Psi) \right]^{2\beta} \rho(\Psi) \right),
\label{eqn:generalized}     
\end{equation}
% where
% \begin{equation}
%      C_\beta^{-1} = \frac{2\pi^{5/2}(1/2-\beta)(-\beta)!}% % {2^{\beta-1/2} \sin \pi(\beta-1/2) (1/2-\beta)!},
% \end{equation}
% and 
where $r(\Psi)$ is the radius at which the potential has a value $\Psi$.

We use the Hernquist density profile $\rho(r) = \rho_s (r/r_s)^{-1}(1+r/r_s)^{-3}$ \cite{Hernquist:1990be}, which shares the $1/r$ density cusp with the NFW profile \cite{Navarro:1996gj} but has finite mass and thus sidesteps issues associated with truncation of the NFW profile.\footnote{Note that the code at github.com/NSphere-SIDM/NSphere-SIDM has been updated to support anisotropic NFW, Hernquist, and Cored profiles and allows loading arbitrary anisotropic initial conditions for any density profile.} It also has, for $\beta \leq 1/2$, an analytic form for anisotropic DFs \cite{Baes:2002tw,AnEvans},
\begin{equation}
  f_1({\cal E}) \propto \tildecalE^{5/2-\beta}{}_2F_1(5-2\beta,1-2\beta,7/2-\beta,\tildecalE),
\end{equation}
with $\tildecalE= {\cal E}a/(GM)$, in terms of the hypergeometric function.

The OM models are isotropic at $r\to0$, have an anisotropy $\beta(r) = r^2/(r^2+r_a^2)$ that increases with radius and approach purely radial orbits ($\beta\to1$) at $r\gg r_a^2$, where $r_a$ is the anisotropy radius.  We also define the parameter $\beta_a = \beta(r_s)$ to be the characteristic anisotropy at the scale radius $r_s$. The DF is obtained by writing $F({\cal E},L) =f(Q)$ with $Q={\cal E} - L^2/(2r_a^2) = \Psi -(1/2)v^2\left[ 1+(r/r_a)^2(1-\mu^2) \right]$, where $\mu=\cos\theta$ is the cosine of the angle the velocity makes with respect to the radial direction.  The quantity $Q$ is a conserved integral of motion of individual orbits in these stationary distributions. Here
\begin{equation}
    f(Q) = \frac{1}{\sqrt{8} \pi^2 m_p} \frac{d}{dQ} \int_0^Q\, \frac{d\Psi}{\sqrt{Q-\Psi}} \frac{d}{d\Psi} \left(\left[1+\frac{r^2}{r_a^2} \right] \rho\right).
\end{equation}
With substitutions ${\cal E} \to Q$ and $\rho \to \rho_Q(r) = (1+r^2/r_a^2)\rho(r)$, Eq.~(\ref{eqn:eddington}) becomes identical to Eq.~(4), and we can sample analogously to the isotropic case \cite{Kamionkowski:2025uae}.  Defining pseudo-velocities $w_r = v_r$ and $w_t = \sqrt{1+r^2/r_a^2}\,v_t$ such that $Q = \Psi - w^2/2$, we sample $w$ from $w^2 F(\Psi - w^2/2)$ and $\mu_w$ uniformly to get $w_r$ and $w_t$, then transform back to physical velocities and angles via $v_r = w_r$, $v_t = w_t/\sqrt{1+r^2/r_a^2}$, $v = \sqrt{v_r^2 + v_t^2}$, and $\mu = v_r/v$.  This transformation automatically generates the anisotropic velocity distribution.  We note OM models only exist for the Hernquist profile for $r_a \geq 0.202\,r_s$, or $\beta_a \leq 0.96$.

We have generated initial conditions for simulations with $N=10^5$ particles for constant-anisotropy models with $\beta=-1/2$, $-1/4$, 0, 1/4, and 1/2.  In each case, we have taken $r_s =1.18$ kpc and $\rho_s= 2.73\times 10^7\, m_\odot~{\rm kpc}^{-3}$ as in Refs.~\cite{Palubski:2024ibb,Kamionkowski:2025fat} (although those were for NFW, rather than Hernquist, profiles), or equivalently, a total mass $M=2.82\times 10^8\, m_\odot$.  The SIDM cross section is isotropic and velocity-independent and has a magnitude (per unit mass) $\sigma/m = 50\, {\rm cm}^2~{\rm g}^{-1}$.  The scattering is implemented as described in Ref.~\cite{Kamionkowski:2025fat} using the code at {\tt github.com/NSphere-SIDM/NSphere-SIDM}, and we now have extended that code to run simulations with several anisotropic DFs.

\begin{figure}
\includegraphics[width=\columnwidth]{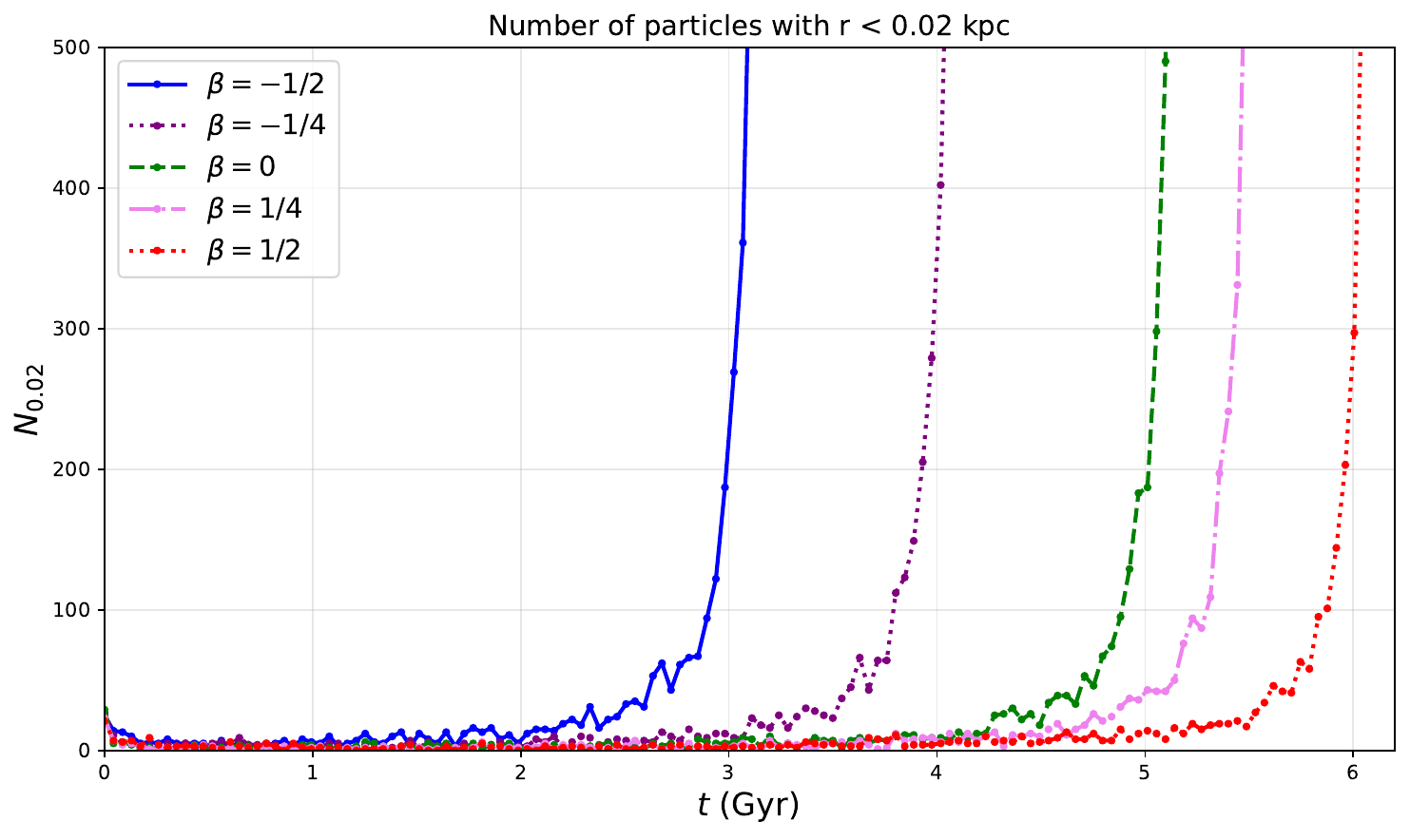}
\includegraphics[width=\columnwidth]{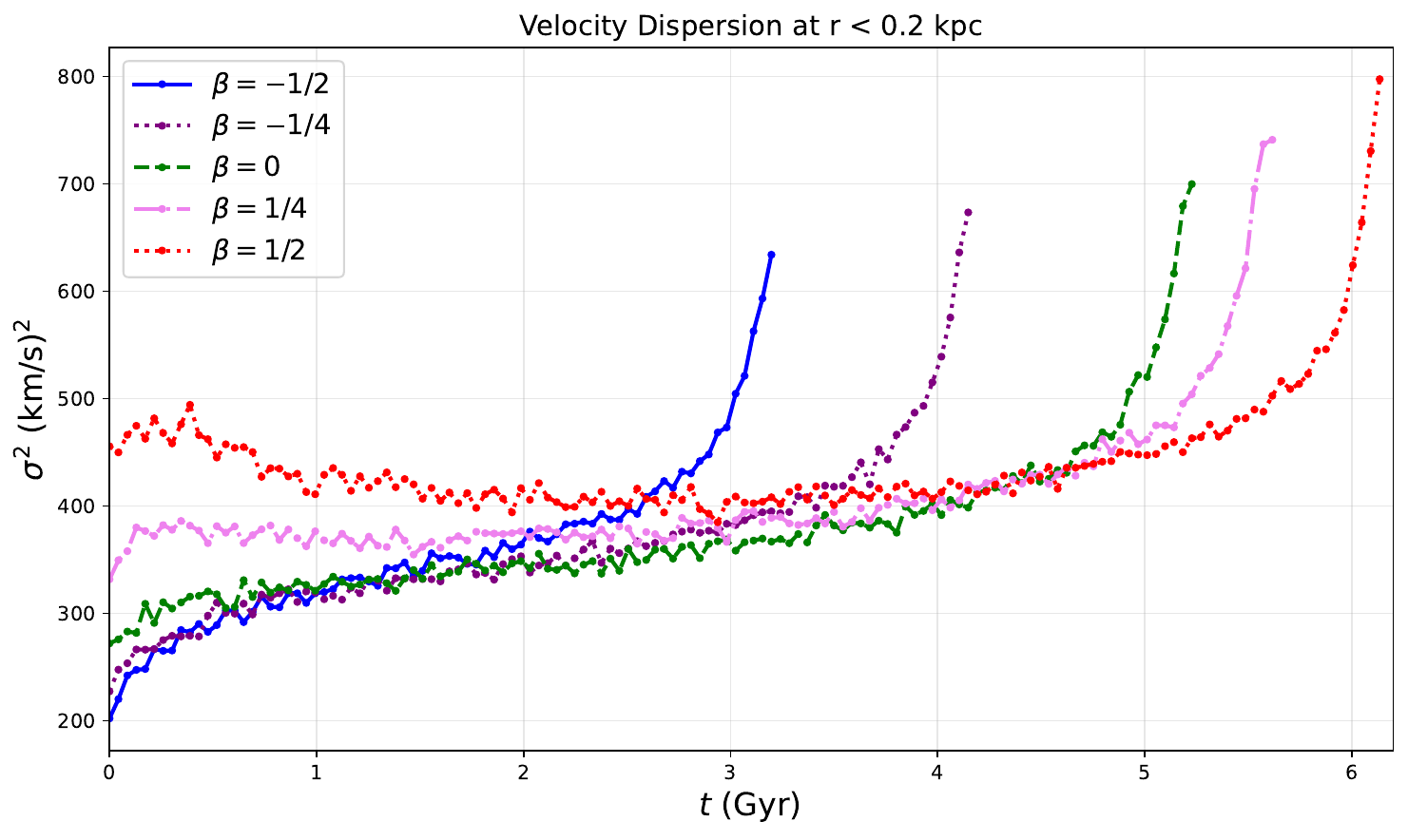}
\includegraphics[width=\columnwidth]{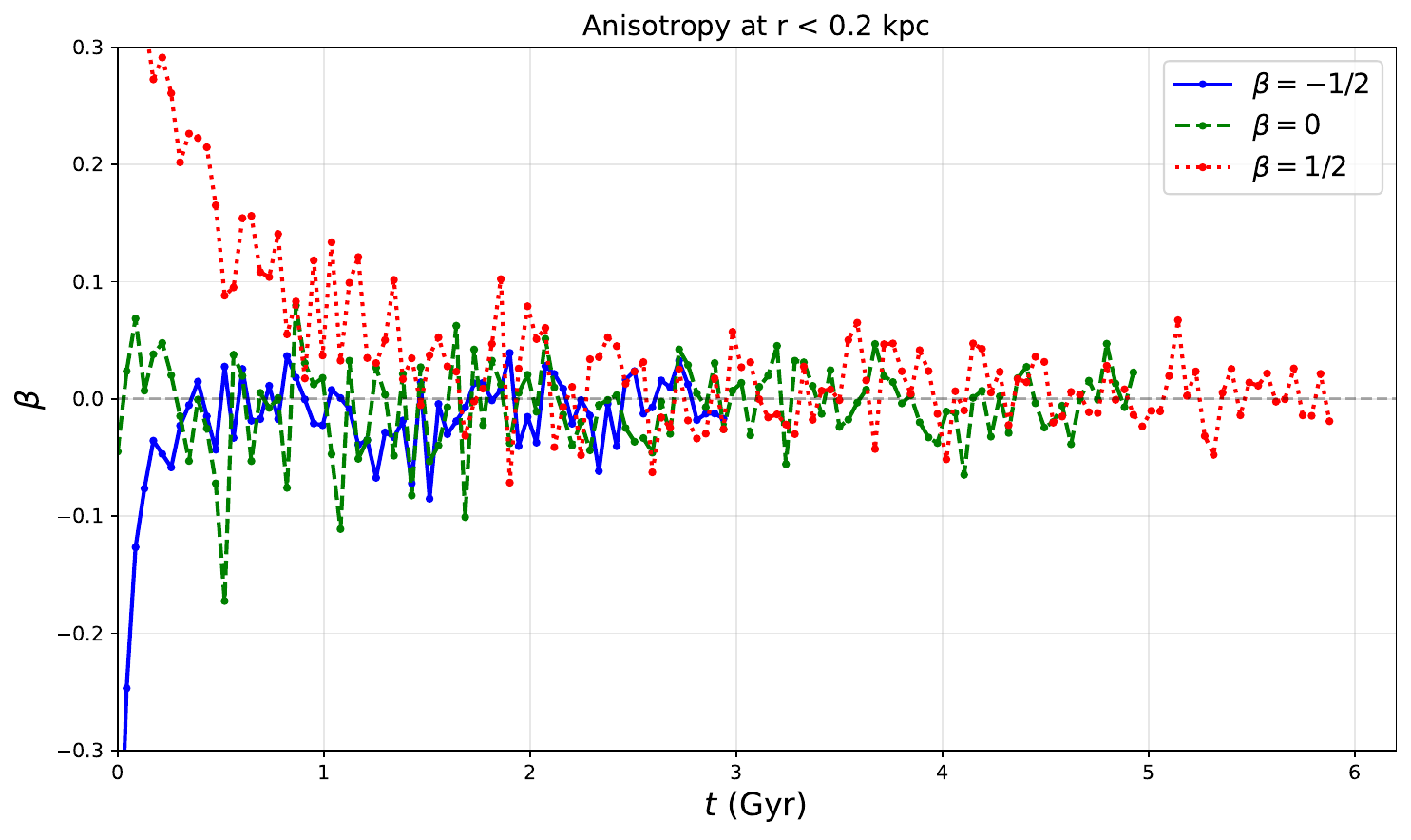}
     \caption{{\it Top:} The evolution of the number of particles with radii $r<0.02$ kpc as a function of time for the five constant-$\beta$ simulations showing collapse time variation between $\sim$3 Gyr and $\sim$6 Gyr.  {\it Middle:}  The evolution of the velocity dispersion at radii $r<0.2$ kpc.  {\it Bottom:} Evolution of the anisotropy parameter $\beta$ (averaged over the innermost 0.2 kpc).}
\label{fig:velocitydispersion}
\end{figure}

\begin{figure}
\includegraphics[width=\columnwidth]{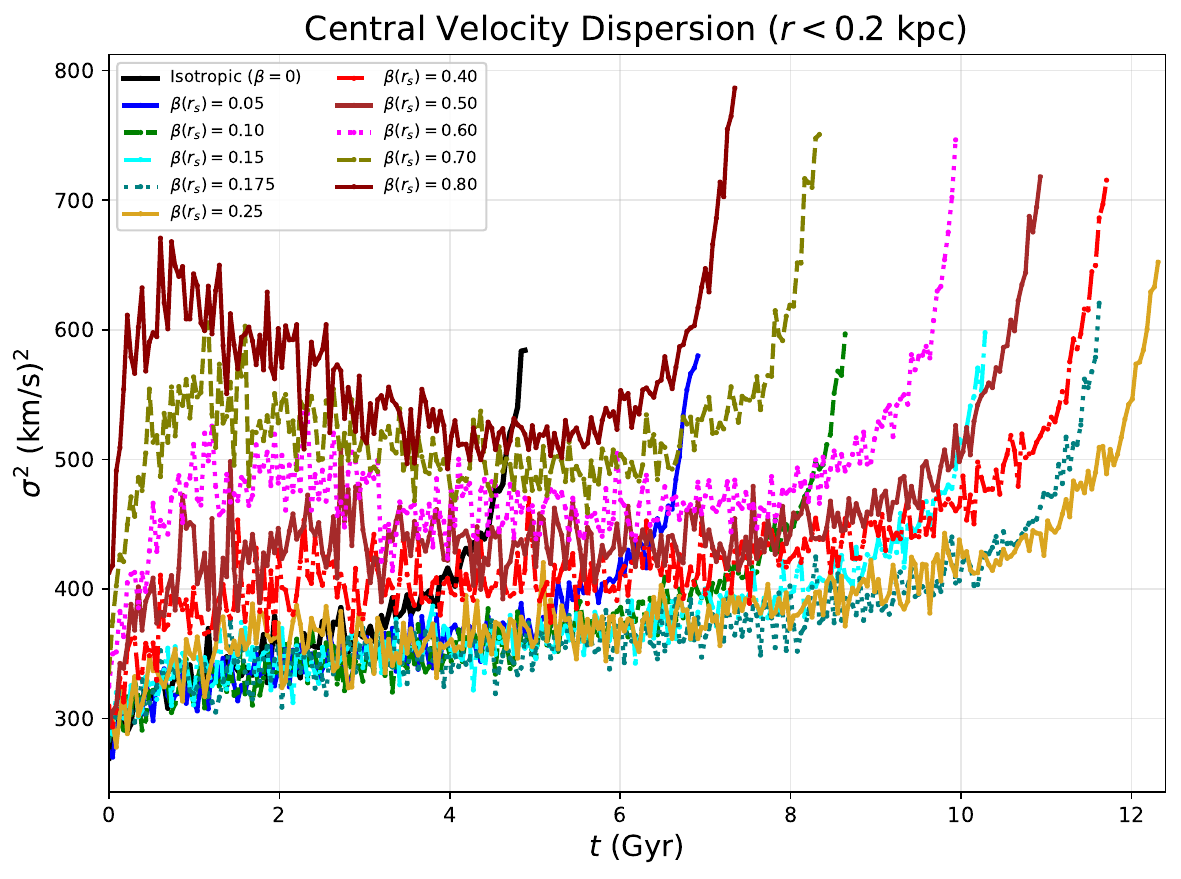}
\includegraphics[width=\columnwidth]{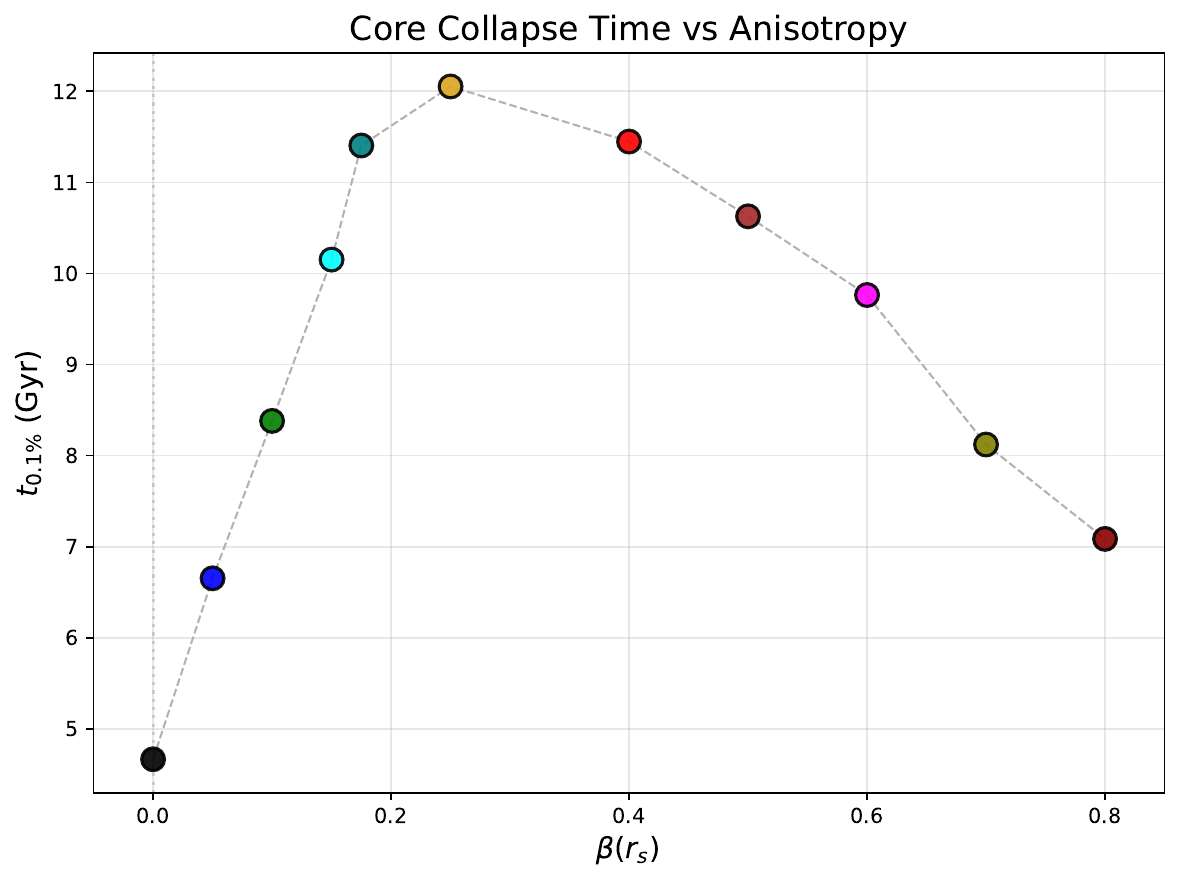}
     \caption{{\it Top:} The evolution of the velocity dispersion at radii $r<0.2$ kpc for OM models.  {\it Bottom:} The collapse times for simulations with different values of initial anisotropy $\beta_a$ at the scale radius $r_s$.}
\label{fig:omresults}
\end{figure}

Fig. \ref{fig:phasespace} shows the initial phase-space distributions for the $\beta=0,\pm1/2$ simulations in the radius--radial-velocity plane.  The three distributions are the same in radius but become increasingly spread out in radial velocity as $\beta$ increases from $-1/2$ (tangential orbits preferred) to 1/2 (radial orbits preferred).  The evolution of the number of particles in the innermost 0.02 kpc is shown for all five simulations in the top panel of Fig.~\ref{fig:velocitydispersion} and that of the velocity dispersion in the innermost kpc in the middle panel.  We also show in the bottom panel the evolution of an anisotropy parameter, defined here in terms of ratios of radial and tangential velocity dispersion averaged over the inner 0.2 kpc (only for $\beta=0,\pm1/2$ to avoid clutter in the Figure). The simulations are halted when the central density becomes so high that our finite time step can no longer resolve the scattering dynamics. The middle panel shows just how dramatically the velocity dispersion can change for different self-consistent halos with the same density profile but different velocity distributions.  We caution that, while collapse times should be physical, our simulation breaks down after collapse onset and the detailed peak values of the velocity dispersion here and in Fig.~\ref{fig:omresults} may not be accurate.  The bottom panel also suggests that the velocity distribution with $\beta=1/2$ takes a bit more time to thermalize than with $\beta=-1/2$. The top panel of Fig.~\ref{fig:velocitydispersion} provides the central result that the gravothermal collapse time can vary from $\sim$3 Gyr to $\sim$6 Gyr, as $\beta$ varies from $-1/2$ to $1/2$.    % We note that the gravothermal collapse times seem to vary (by a few percent for $10^5$ particles) from one simulation to the next \cite{Kamionkowski:2025fat}.  
% Even so, the results shown in the middle panel of Fig.~\ref{fig:velocitydispersion} indicate a trend for increasing collapse timescale with increasing $\beta$.

\begin{figure}
\includegraphics[width=\columnwidth]{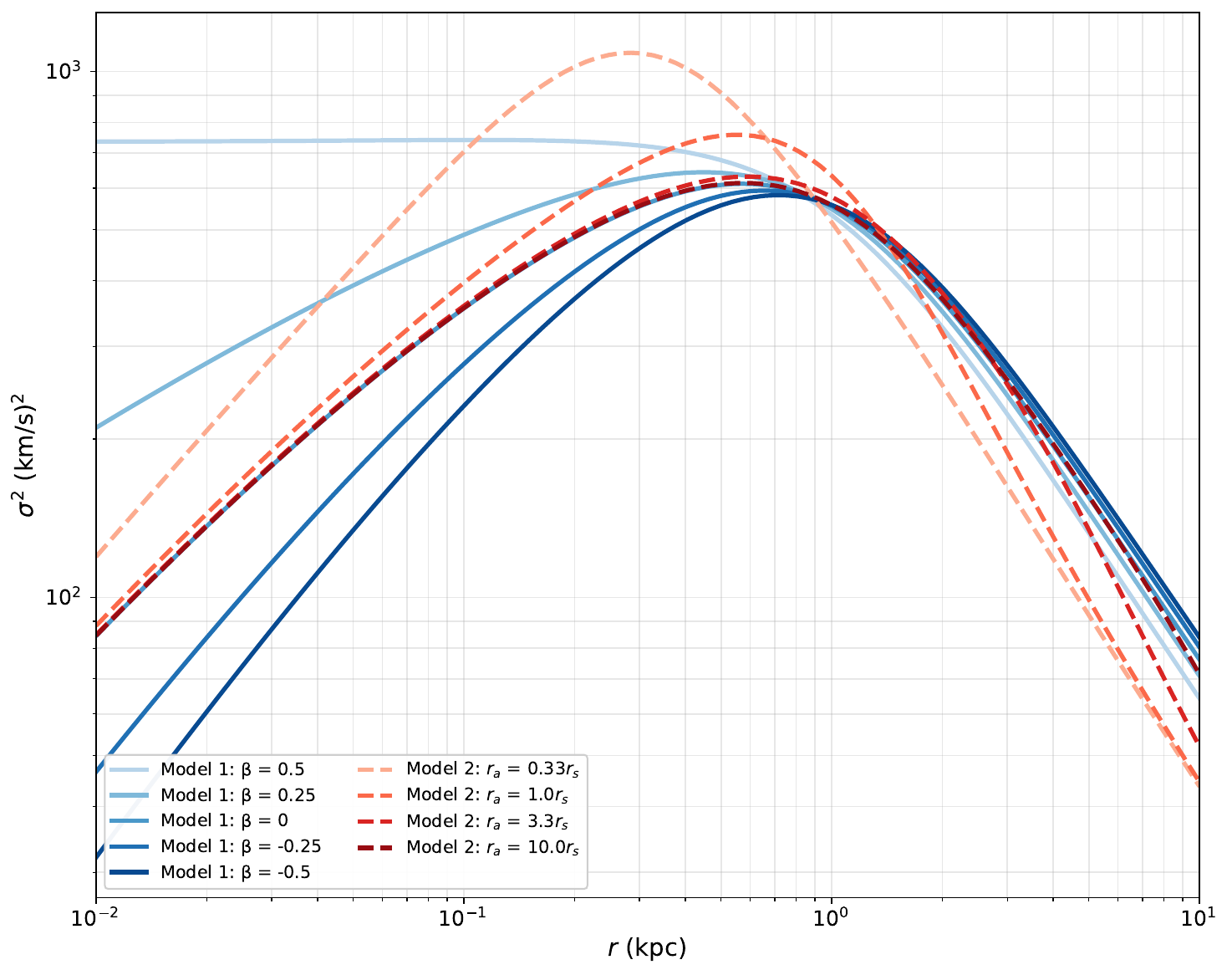}
     \caption{The initial velocity dispersion as a function of radius for 5 models with constant $\beta$ and 5 OM models.}
\label{fig:dispersionprofile}
\end{figure}

The results of a suite of simulations of OM models with different values of $r_a$ (parameterized here in terms of anisotropy $\beta_a$ at the scale radius) are shown in Fig.~\ref{fig:omresults}.  It may come as somewhat of a surprise that the collapse time does not vary monotonically with $\beta_a$, but this can be understood in terms of the initial velocity-dispersion profiles, as we explain below.

The radial velocity dispersion can be obtained as a function of radius through the Jeans theorem (e.g., Ref.~\cite{Binney:2008},
\begin{equation}
     \frac{d(\rho \vev{v_r^2})}{dr} + 2 \frac{\beta}{r} \rho \vev{v_r^2} = -\rho \frac{d\Phi}{dr},
\label{eqn:jeans}     
\end{equation}
and then the three-dimensional velocity dispersion profile is
$\vev{v^2} = \vev{v_r^2}\left[1+2 \beta\right]$.
These dispersions are shown in Fig.~\ref{fig:dispersionprofile} for both constant- and increasing-$\beta$ models.  The constant-$\beta$ dispersions can be understood as follows:  When the orbits are highly radial, the origin sees particles that have orbits that extend to large radii.  In the limit that orbits are all tangential, the orbits are circular with velocities dependent on the enclosed mass; i.e., $v^2 \propto r$ as $r\to0$.  

The dependence of the collapse timescale on $\beta$ in the constant-$\beta$ models can most likely be understood in terms of the heat transport that SIDM introduces and thus the {\it gradient} of the velocity dispersion.\footnote{The gradient drives heat flux ${\bf Q} \propto -\kappa \partial \langle v^2 \rangle/\partial r \, \hat{\bf r}$ in the gravothermal fluid approximation (e.g., \cite{Koda:2011yb,Yang:2022zkd}).}  This is smallest for $\beta=1/2$ (longest collapse time) and largest for $\beta=-1/2$ (shortest time), suggesting that redistribution of heat within the halo may be quicker for the colder interiors associated with tangential orbits, something seen also in the middle panel of Fig.~\ref{fig:velocitydispersion}.

The decrease of collapse timescale with increasing $\beta_a$ at large $\beta_a$ in the OM models is more difficult to explain heuristically.  The change from increasing collapse time to decreasing at larger $\beta_a$ traces to a very subtle effect.  Animations of the phase-space-distribution evolution for all of the constant-$\beta$ simulations and OM simulations with smaller $\beta_a$, generated with our public code, show a smooth evolution of the phase-space density.  However, for larger values of $\beta_a$ there is a noticeable outflow of particles from the center, stronger with larger $\beta_a$, as soon as the simulation is started.  This is a consequence of the large velocity dispersions that these models have at small radii combined with very large initial scattering rates at small radii.  We surmise that the subsequent evolution thus depends on some complex interplay of substructure in the evolving phase-space distribution that cannot be understood in terms of smooth evolution to a roughly isothermal state.

This observation does, though, reinforce an important point.  It is clear that the initial conditions---a self-gravitating halo with a velocity distribution consistent with collisionless particles---in such simulations are not always realistic.  They may be a good starting point in the limit of sufficiently weak SIDM interactions but as the SIDM interaction rate is increased the likelihood that such initial conditions may approximate those in a realistic halo decrease.  Conclusions about collapse timescales based on simulations with self-consistent collisionless initial conditions become increasingly unreliable.  One alternative may perhaps be to run a cosmological simulation with SIDM and then identify from that a DF that approximates the one to which SIDM dark matter may initially virialize.  A simulation could then be initialized with these DFs to more robustly determine collapse timescales.

To recapitulate, specification of a density profile for a self-gravitating system does not uniquely determine its velocity distribution.  Here we have shown with some simple numerical experiments that the timescale for gravothermal collapse of an SIDM halo with an initial $1/r$ density cusp can vary by as much as a factor of $\sim$2 with reasonable variation in the direction dependence of the velocity distribution.  We did so with constant-anisotropy models and with models that become increasingly anisotropic (and radial) at larger radii.  

The dynamics of spherical infall suggest that orbits should be preferentially radial at larger radii, while symmetry dictates that the DF should become isotropic near the center, as in the OM case.  Still, as discussed earlier one may not immediately conclude that these models are more realistic than the constant-$\beta$ models.  One analysis \cite{Lemze_2012} of simulated halos finds isotropic orbits near the center that then increase to anisotropy, but only to $\beta\sim0.3$ at $r\gtrsim0.2\, r_s$.  Another \cite{Wojtak:2013eia} finds larger anisotropy, $\beta\sim0.4$ at all radii, more like the constant-anisotropy models, although detailed results may depend on halo parameters \cite{He:2024gvw}.  Observations find agreement with the expectation of isotropy at the center and more radial orbits further out \cite{Biviano:2013eia}.  The takeaway is that anisotropy and halo-to-halo variation in anisotropy should be considered an essential ingredient in future modeling of gravothermal collapse and in efforts to connect SIDM models with observations.  This variation may be comparable to that from the SIDM cross-section itself.

\medskip
We thank James Gurian for useful discussions.  This work was supported at JHU by NSF Grant No.\ 2412361, NASA ATP Grant No.\ 80NSSC24K1226, and the Templeton Foundation and at UBC by a NSERC of Canada Discovery Grant.

\bibliography{paper}% Produces the bibliography via BibTeX.

\begin{thebibliography}{46}%
\makeatletter
\providecommand \@ifxundefined [1]{%
 \@ifx{#1\undefined}
}%
\providecommand \@ifnum [1]{%
 \ifnum #1\expandafter \@firstoftwo
 \else \expandafter \@secondoftwo
 \fi
}%
\providecommand \@ifx [1]{%
 \ifx #1\expandafter \@firstoftwo
 \else \expandafter \@secondoftwo
 \fi
}%
\providecommand \natexlab [1]{#1}%
\providecommand \enquote  [1]{``#1''}%
\providecommand \bibnamefont  [1]{#1}%
\providecommand \bibfnamefont [1]{#1}%
\providecommand \citenamefont [1]{#1}%
\providecommand \href@noop [0]{\@secondoftwo}%
\providecommand \href [0]{\begingroup \@sanitize@url \@href}%
\providecommand \@href[1]{\@@startlink{#1}\@@href}%
\providecommand \@@href[1]{\endgroup#1\@@endlink}%
\providecommand \@sanitize@url [0]{\catcode `\\12\catcode `\$12\catcode
  `\&12\catcode `\#12\catcode `\^12\catcode `\_12\catcode `\%12\relax}%
\providecommand \@@startlink[1]{}%
\providecommand \@@endlink[0]{}%
\providecommand \url  [0]{\begingroup\@sanitize@url \@url }%
\providecommand \@url [1]{\endgroup\@href {#1}{\urlprefix }}%
\providecommand \urlprefix  [0]{URL }%
\providecommand \Eprint [0]{\href }%
\providecommand \doibase [0]{https://doi.org/}%
\providecommand \selectlanguage [0]{\@gobble}%
\providecommand \bibinfo  [0]{\@secondoftwo}%
\providecommand \bibfield  [0]{\@secondoftwo}%
\providecommand \translation [1]{[#1]}%
\providecommand \BibitemOpen [0]{}%
\providecommand \bibitemStop [0]{}%
\providecommand \bibitemNoStop [0]{.\EOS\space}%
\providecommand \EOS [0]{\spacefactor3000\relax}%
\providecommand \BibitemShut  [1]{\csname bibitem#1\endcsname}%
\let\auto@bib@innerbib\@empty
%</preamble>
\bibitem [{\citenamefont {Tulin}\ and\ \citenamefont
  {Yu}(2018)}]{Tulin:2017ara}%
  \BibitemOpen
  \bibfield  {author} {\bibinfo {author} {\bibfnamefont {S.}~\bibnamefont
  {Tulin}}\ and\ \bibinfo {author} {\bibfnamefont {H.-B.}\ \bibnamefont {Yu}},\
  }\bibfield  {title} {\bibinfo {title} {{Dark Matter Self-interactions and
  Small Scale Structure}},\ }\href
  {https://doi.org/10.1016/j.physrep.2017.11.004} {\bibfield  {journal}
  {\bibinfo  {journal} {Phys. Rept.}\ }\textbf {\bibinfo {volume} {730}},\
  \bibinfo {pages} {1} (\bibinfo {year} {2018})},\ \Eprint
  {https://arxiv.org/abs/1705.02358} {arXiv:1705.02358 [hep-ph]} \BibitemShut
  {NoStop}%
\bibitem [{\citenamefont {Buckley}\ and\ \citenamefont
  {Peter}(2018)}]{Buckley:2017ijx}%
  \BibitemOpen
  \bibfield  {author} {\bibinfo {author} {\bibfnamefont {M.~R.}\ \bibnamefont
  {Buckley}}\ and\ \bibinfo {author} {\bibfnamefont {A.~H.~G.}\ \bibnamefont
  {Peter}},\ }\bibfield  {title} {\bibinfo {title} {{Gravitational probes of
  dark matter physics}},\ }\href
  {https://doi.org/10.1016/j.physrep.2018.07.003} {\bibfield  {journal}
  {\bibinfo  {journal} {Phys. Rept.}\ }\textbf {\bibinfo {volume} {761}},\
  \bibinfo {pages} {1} (\bibinfo {year} {2018})},\ \Eprint
  {https://arxiv.org/abs/1712.06615} {arXiv:1712.06615 [astro-ph.CO]}
  \BibitemShut {NoStop}%
\bibitem [{\citenamefont {Spergel}\ and\ \citenamefont
  {Steinhardt}(2000)}]{Spergel:1999mh}%
  \BibitemOpen
  \bibfield  {author} {\bibinfo {author} {\bibfnamefont {D.~N.}\ \bibnamefont
  {Spergel}}\ and\ \bibinfo {author} {\bibfnamefont {P.~J.}\ \bibnamefont
  {Steinhardt}},\ }\bibfield  {title} {\bibinfo {title} {{Observational
  evidence for selfinteracting cold dark matter}},\ }\href
  {https://doi.org/10.1103/PhysRevLett.84.3760} {\bibfield  {journal} {\bibinfo
   {journal} {Phys. Rev. Lett.}\ }\textbf {\bibinfo {volume} {84}},\ \bibinfo
  {pages} {3760} (\bibinfo {year} {2000})},\ \Eprint
  {https://arxiv.org/abs/astro-ph/9909386} {arXiv:astro-ph/9909386}
  \BibitemShut {NoStop}%
\bibitem [{\citenamefont {Kaplinghat}\ \emph {et~al.}(2016)\citenamefont
  {Kaplinghat}, \citenamefont {Tulin},\ and\ \citenamefont
  {Yu}}]{Kaplinghat:2015aga}%
  \BibitemOpen
  \bibfield  {author} {\bibinfo {author} {\bibfnamefont {M.}~\bibnamefont
  {Kaplinghat}}, \bibinfo {author} {\bibfnamefont {S.}~\bibnamefont {Tulin}},\
  and\ \bibinfo {author} {\bibfnamefont {H.-B.}\ \bibnamefont {Yu}},\
  }\bibfield  {title} {\bibinfo {title} {{Dark Matter Halos as Particle
  Colliders: Unified Solution to Small-Scale Structure Puzzles from Dwarfs to
  Clusters}},\ }\href {https://doi.org/10.1103/PhysRevLett.116.041302}
  {\bibfield  {journal} {\bibinfo  {journal} {Phys. Rev. Lett.}\ }\textbf
  {\bibinfo {volume} {116}},\ \bibinfo {pages} {041302} (\bibinfo {year}
  {2016})},\ \Eprint {https://arxiv.org/abs/1508.03339} {arXiv:1508.03339
  [astro-ph.CO]} \BibitemShut {NoStop}%
\bibitem [{\citenamefont {Kamada}\ \emph {et~al.}(2017)\citenamefont {Kamada},
  \citenamefont {Kaplinghat}, \citenamefont {Pace},\ and\ \citenamefont
  {Yu}}]{Kamada:2016euw}%
  \BibitemOpen
  \bibfield  {author} {\bibinfo {author} {\bibfnamefont {A.}~\bibnamefont
  {Kamada}}, \bibinfo {author} {\bibfnamefont {M.}~\bibnamefont {Kaplinghat}},
  \bibinfo {author} {\bibfnamefont {A.~B.}\ \bibnamefont {Pace}},\ and\
  \bibinfo {author} {\bibfnamefont {H.-B.}\ \bibnamefont {Yu}},\ }\bibfield
  {title} {\bibinfo {title} {{How the Self-Interacting Dark Matter Model
  Explains the Diverse Galactic Rotation Curves}},\ }\href
  {https://doi.org/10.1103/PhysRevLett.119.111102} {\bibfield  {journal}
  {\bibinfo  {journal} {Phys. Rev. Lett.}\ }\textbf {\bibinfo {volume} {119}},\
  \bibinfo {pages} {111102} (\bibinfo {year} {2017})},\ \Eprint
  {https://arxiv.org/abs/1611.02716} {arXiv:1611.02716 [astro-ph.GA]}
  \BibitemShut {NoStop}%
\bibitem [{\citenamefont {Correa}(2021)}]{Correa:2020qam}%
  \BibitemOpen
  \bibfield  {author} {\bibinfo {author} {\bibfnamefont {C.~A.}\ \bibnamefont
  {Correa}},\ }\bibfield  {title} {\bibinfo {title} {{Constraining
  velocity-dependent self-interacting dark matter with the Milky
  Way\textquoteright{}s dwarf spheroidal galaxies}},\ }\href
  {https://doi.org/10.1093/mnras/stab506} {\bibfield  {journal} {\bibinfo
  {journal} {Mon. Not. Roy. Astron. Soc.}\ }\textbf {\bibinfo {volume} {503}},\
  \bibinfo {pages} {920} (\bibinfo {year} {2021})},\ \Eprint
  {https://arxiv.org/abs/2007.02958} {arXiv:2007.02958 [astro-ph.GA]}
  \BibitemShut {NoStop}%
\bibitem [{\citenamefont {Zeng}\ \emph {et~al.}(2022)\citenamefont {Zeng},
  \citenamefont {Peter}, \citenamefont {Du}, \citenamefont {Benson},
  \citenamefont {Kim}, \citenamefont {Jiang}, \citenamefont {Cyr-Racine},\ and\
  \citenamefont {Vogelsberger}}]{Zeng:2021ldo}%
  \BibitemOpen
  \bibfield  {author} {\bibinfo {author} {\bibfnamefont {Z.~C.}\ \bibnamefont
  {Zeng}}, \bibinfo {author} {\bibfnamefont {A.~H.~G.}\ \bibnamefont {Peter}},
  \bibinfo {author} {\bibfnamefont {X.}~\bibnamefont {Du}}, \bibinfo {author}
  {\bibfnamefont {A.}~\bibnamefont {Benson}}, \bibinfo {author} {\bibfnamefont
  {S.}~\bibnamefont {Kim}}, \bibinfo {author} {\bibfnamefont {F.}~\bibnamefont
  {Jiang}}, \bibinfo {author} {\bibfnamefont {F.-Y.}\ \bibnamefont
  {Cyr-Racine}},\ and\ \bibinfo {author} {\bibfnamefont {M.}~\bibnamefont
  {Vogelsberger}},\ }\bibfield  {title} {\bibinfo {title} {{Core-collapse,
  evaporation, and tidal effects: the life story of a self-interacting dark
  matter subhalo}},\ }\href {https://doi.org/10.1093/mnras/stac1094} {\bibfield
   {journal} {\bibinfo  {journal} {Mon. Not. Roy. Astron. Soc.}\ }\textbf
  {\bibinfo {volume} {513}},\ \bibinfo {pages} {4845} (\bibinfo {year}
  {2022})},\ \Eprint {https://arxiv.org/abs/2110.00259} {arXiv:2110.00259
  [astro-ph.CO]} \BibitemShut {NoStop}%
\bibitem [{\citenamefont {Zentner}\ \emph {et~al.}(2022)\citenamefont
  {Zentner}, \citenamefont {Dandavate}, \citenamefont {Slone},\ and\
  \citenamefont {Lisanti}}]{Zentner:2022xux}%
  \BibitemOpen
  \bibfield  {author} {\bibinfo {author} {\bibfnamefont {A.}~\bibnamefont
  {Zentner}}, \bibinfo {author} {\bibfnamefont {S.}~\bibnamefont {Dandavate}},
  \bibinfo {author} {\bibfnamefont {O.}~\bibnamefont {Slone}},\ and\ \bibinfo
  {author} {\bibfnamefont {M.}~\bibnamefont {Lisanti}},\ }\bibfield  {title}
  {\bibinfo {title} {{A critical assessment of solutions to the galaxy
  diversity problem}},\ }\href {https://doi.org/10.1088/1475-7516/2022/07/031}
  {\bibfield  {journal} {\bibinfo  {journal} {JCAP}\ }\textbf {\bibinfo
  {volume} {07}}\bibfield  {number} {\bibinfo  {number} { (07)},\ \bibinfo
  {pages} {031}},\ }\Eprint {https://arxiv.org/abs/2202.00012}
  {arXiv:2202.00012 [astro-ph.GA]} \BibitemShut {NoStop}%
\bibitem [{\citenamefont {Correa}\ \emph {et~al.}(2022)\citenamefont {Correa},
  \citenamefont {Schaller}, \citenamefont {Ploeckinger}, \citenamefont
  {Anau~Montel}, \citenamefont {Weniger},\ and\ \citenamefont
  {Ando}}]{Correa:2022dey}%
  \BibitemOpen
  \bibfield  {author} {\bibinfo {author} {\bibfnamefont {C.~A.}\ \bibnamefont
  {Correa}}, \bibinfo {author} {\bibfnamefont {M.}~\bibnamefont {Schaller}},
  \bibinfo {author} {\bibfnamefont {S.}~\bibnamefont {Ploeckinger}}, \bibinfo
  {author} {\bibfnamefont {N.}~\bibnamefont {Anau~Montel}}, \bibinfo {author}
  {\bibfnamefont {C.}~\bibnamefont {Weniger}},\ and\ \bibinfo {author}
  {\bibfnamefont {S.}~\bibnamefont {Ando}},\ }\bibfield  {title} {\bibinfo
  {title} {{TangoSIDM: tantalizing models of self-interacting dark matter}},\
  }\href {https://doi.org/10.1093/mnras/stac2830} {\bibfield  {journal}
  {\bibinfo  {journal} {Mon. Not. Roy. Astron. Soc.}\ }\textbf {\bibinfo
  {volume} {517}},\ \bibinfo {pages} {3045} (\bibinfo {year} {2022})},\ \Eprint
  {https://arxiv.org/abs/2206.11298} {arXiv:2206.11298 [astro-ph.GA]}
  \BibitemShut {NoStop}%
\bibitem [{\citenamefont {Yang}\ \emph
  {et~al.}(2023{\natexlab{a}})\citenamefont {Yang}, \citenamefont {Nadler},\
  and\ \citenamefont {Yu}}]{Yang:2022mxl}%
  \BibitemOpen
  \bibfield  {author} {\bibinfo {author} {\bibfnamefont {D.}~\bibnamefont
  {Yang}}, \bibinfo {author} {\bibfnamefont {E.~O.}\ \bibnamefont {Nadler}},\
  and\ \bibinfo {author} {\bibfnamefont {H.-B.}\ \bibnamefont {Yu}},\
  }\bibfield  {title} {\bibinfo {title} {{Strong Dark Matter Self-interactions
  Diversify Halo Populations within and surrounding the Milky Way}},\ }\href
  {https://doi.org/10.3847/1538-4357/acc73e} {\bibfield  {journal} {\bibinfo
  {journal} {Astrophys. J.}\ }\textbf {\bibinfo {volume} {949}},\ \bibinfo
  {pages} {67} (\bibinfo {year} {2023}{\natexlab{a}})},\ \Eprint
  {https://arxiv.org/abs/2211.13768} {arXiv:2211.13768 [astro-ph.GA]}
  \BibitemShut {NoStop}%
\bibitem [{\citenamefont {Nadler}\ \emph {et~al.}(2023)\citenamefont {Nadler},
  \citenamefont {Yang},\ and\ \citenamefont {Yu}}]{Nadler:2023nrd}%
  \BibitemOpen
  \bibfield  {author} {\bibinfo {author} {\bibfnamefont {E.~O.}\ \bibnamefont
  {Nadler}}, \bibinfo {author} {\bibfnamefont {D.}~\bibnamefont {Yang}},\ and\
  \bibinfo {author} {\bibfnamefont {H.-B.}\ \bibnamefont {Yu}},\ }\bibfield
  {title} {\bibinfo {title} {{A Self-interacting Dark Matter Solution to the
  Extreme Diversity of Low-mass Halo Properties}},\ }\href
  {https://doi.org/10.3847/2041-8213/ad0e09} {\bibfield  {journal} {\bibinfo
  {journal} {Astrophys. J. Lett.}\ }\textbf {\bibinfo {volume} {958}},\
  \bibinfo {pages} {L39} (\bibinfo {year} {2023})},\ \Eprint
  {https://arxiv.org/abs/2306.01830} {arXiv:2306.01830 [astro-ph.GA]}
  \BibitemShut {NoStop}%
\bibitem [{\citenamefont {Colin}\ \emph {et~al.}(2002)\citenamefont {Colin},
  \citenamefont {Avila-Reese}, \citenamefont {Valenzuela},\ and\ \citenamefont
  {Firmani}}]{Colin:2002nk}%
  \BibitemOpen
  \bibfield  {author} {\bibinfo {author} {\bibfnamefont {P.}~\bibnamefont
  {Colin}}, \bibinfo {author} {\bibfnamefont {V.}~\bibnamefont {Avila-Reese}},
  \bibinfo {author} {\bibfnamefont {O.}~\bibnamefont {Valenzuela}},\ and\
  \bibinfo {author} {\bibfnamefont {C.}~\bibnamefont {Firmani}},\ }\bibfield
  {title} {\bibinfo {title} {{Structure and subhalo population of halos in a
  selfinteracting dark matter cosmology}},\ }\href
  {https://doi.org/10.1086/344259} {\bibfield  {journal} {\bibinfo  {journal}
  {Astrophys. J.}\ }\textbf {\bibinfo {volume} {581}},\ \bibinfo {pages} {777}
  (\bibinfo {year} {2002})},\ \Eprint {https://arxiv.org/abs/astro-ph/0205322}
  {arXiv:astro-ph/0205322} \BibitemShut {NoStop}%
\bibitem [{\citenamefont {Kochanek}\ and\ \citenamefont
  {White}(2000)}]{Kochanek:2000pi}%
  \BibitemOpen
  \bibfield  {author} {\bibinfo {author} {\bibfnamefont {C.~S.}\ \bibnamefont
  {Kochanek}}\ and\ \bibinfo {author} {\bibfnamefont {M.~J.}\ \bibnamefont
  {White}},\ }\bibfield  {title} {\bibinfo {title} {{A Quantitative study of
  interacting dark matter in halos}},\ }\href {https://doi.org/10.1086/317149}
  {\bibfield  {journal} {\bibinfo  {journal} {Astrophys. J.}\ }\textbf
  {\bibinfo {volume} {543}},\ \bibinfo {pages} {514} (\bibinfo {year}
  {2000})},\ \Eprint {https://arxiv.org/abs/astro-ph/0003483}
  {arXiv:astro-ph/0003483} \BibitemShut {NoStop}%
\bibitem [{\citenamefont {Balberg}\ and\ \citenamefont
  {Shapiro}(2002)}]{Balberg:2001qg}%
  \BibitemOpen
  \bibfield  {author} {\bibinfo {author} {\bibfnamefont {S.}~\bibnamefont
  {Balberg}}\ and\ \bibinfo {author} {\bibfnamefont {S.~L.}\ \bibnamefont
  {Shapiro}},\ }\bibfield  {title} {\bibinfo {title} {{Gravothermal collapse of
  selfinteracting dark matter halos and the origin of massive black holes}},\
  }\href {https://doi.org/10.1103/PhysRevLett.88.101301} {\bibfield  {journal}
  {\bibinfo  {journal} {Phys. Rev. Lett.}\ }\textbf {\bibinfo {volume} {88}},\
  \bibinfo {pages} {101301} (\bibinfo {year} {2002})},\ \Eprint
  {https://arxiv.org/abs/astro-ph/0111176} {arXiv:astro-ph/0111176}
  \BibitemShut {NoStop}%
\bibitem [{\citenamefont {Koda}\ and\ \citenamefont
  {Shapiro}(2011)}]{Koda:2011yb}%
  \BibitemOpen
  \bibfield  {author} {\bibinfo {author} {\bibfnamefont {J.}~\bibnamefont
  {Koda}}\ and\ \bibinfo {author} {\bibfnamefont {P.~R.}\ \bibnamefont
  {Shapiro}},\ }\bibfield  {title} {\bibinfo {title} {{Gravothermal collapse of
  isolated self-interacting dark matter haloes: N-body simulation versus the
  fluid model}},\ }\href {https://doi.org/10.1111/j.1365-2966.2011.18684.x}
  {\bibfield  {journal} {\bibinfo  {journal} {Mon. Not. Roy. Astron. Soc.}\
  }\textbf {\bibinfo {volume} {415}},\ \bibinfo {pages} {1125} (\bibinfo {year}
  {2011})},\ \Eprint {https://arxiv.org/abs/1101.3097} {arXiv:1101.3097
  [astro-ph.CO]} \BibitemShut {NoStop}%
\bibitem [{\citenamefont {Essig}\ \emph {et~al.}(2019)\citenamefont {Essig},
  \citenamefont {Mcdermott}, \citenamefont {Yu},\ and\ \citenamefont
  {Zhong}}]{Essig:2018pzq}%
  \BibitemOpen
  \bibfield  {author} {\bibinfo {author} {\bibfnamefont {R.}~\bibnamefont
  {Essig}}, \bibinfo {author} {\bibfnamefont {S.~D.}\ \bibnamefont
  {Mcdermott}}, \bibinfo {author} {\bibfnamefont {H.-B.}\ \bibnamefont {Yu}},\
  and\ \bibinfo {author} {\bibfnamefont {Y.-M.}\ \bibnamefont {Zhong}},\
  }\bibfield  {title} {\bibinfo {title} {{Constraining Dissipative Dark Matter
  Self-Interactions}},\ }\href {https://doi.org/10.1103/PhysRevLett.123.121102}
  {\bibfield  {journal} {\bibinfo  {journal} {Phys. Rev. Lett.}\ }\textbf
  {\bibinfo {volume} {123}},\ \bibinfo {pages} {121102} (\bibinfo {year}
  {2019})},\ \Eprint {https://arxiv.org/abs/1809.01144} {arXiv:1809.01144
  [hep-ph]} \BibitemShut {NoStop}%
\bibitem [{\citenamefont {Nishikawa}\ \emph {et~al.}(2020)\citenamefont
  {Nishikawa}, \citenamefont {Boddy},\ and\ \citenamefont
  {Kaplinghat}}]{Nishikawa:2019lsc}%
  \BibitemOpen
  \bibfield  {author} {\bibinfo {author} {\bibfnamefont {H.}~\bibnamefont
  {Nishikawa}}, \bibinfo {author} {\bibfnamefont {K.~K.}\ \bibnamefont
  {Boddy}},\ and\ \bibinfo {author} {\bibfnamefont {M.}~\bibnamefont
  {Kaplinghat}},\ }\bibfield  {title} {\bibinfo {title} {{Accelerated core
  collapse in tidally stripped self-interacting dark matter halos}},\ }\href
  {https://doi.org/10.1103/PhysRevD.101.063009} {\bibfield  {journal} {\bibinfo
   {journal} {Phys. Rev. D}\ }\textbf {\bibinfo {volume} {101}},\ \bibinfo
  {pages} {063009} (\bibinfo {year} {2020})},\ \Eprint
  {https://arxiv.org/abs/1901.00499} {arXiv:1901.00499 [astro-ph.GA]}
  \BibitemShut {NoStop}%
\bibitem [{\citenamefont {Slone}\ \emph {et~al.}(2023)\citenamefont {Slone},
  \citenamefont {Jiang}, \citenamefont {Lisanti},\ and\ \citenamefont
  {Kaplinghat}}]{Slone:2021nqd}%
  \BibitemOpen
  \bibfield  {author} {\bibinfo {author} {\bibfnamefont {O.}~\bibnamefont
  {Slone}}, \bibinfo {author} {\bibfnamefont {F.}~\bibnamefont {Jiang}},
  \bibinfo {author} {\bibfnamefont {M.}~\bibnamefont {Lisanti}},\ and\ \bibinfo
  {author} {\bibfnamefont {M.}~\bibnamefont {Kaplinghat}},\ }\bibfield  {title}
  {\bibinfo {title} {{Orbital evolution of satellite galaxies in
  self-interacting dark matter models}},\ }\href
  {https://doi.org/10.1103/PhysRevD.107.043014} {\bibfield  {journal} {\bibinfo
   {journal} {Phys. Rev. D}\ }\textbf {\bibinfo {volume} {107}},\ \bibinfo
  {pages} {043014} (\bibinfo {year} {2023})},\ \Eprint
  {https://arxiv.org/abs/2108.03243} {arXiv:2108.03243 [astro-ph.CO]}
  \BibitemShut {NoStop}%
\bibitem [{\citenamefont {Outmezguine}\ \emph {et~al.}(2023)\citenamefont
  {Outmezguine}, \citenamefont {Boddy}, \citenamefont {Gad-Nasr}, \citenamefont
  {Kaplinghat},\ and\ \citenamefont {Sagunski}}]{Outmezguine:2022bhq}%
  \BibitemOpen
  \bibfield  {author} {\bibinfo {author} {\bibfnamefont {N.~J.}\ \bibnamefont
  {Outmezguine}}, \bibinfo {author} {\bibfnamefont {K.~K.}\ \bibnamefont
  {Boddy}}, \bibinfo {author} {\bibfnamefont {S.}~\bibnamefont {Gad-Nasr}},
  \bibinfo {author} {\bibfnamefont {M.}~\bibnamefont {Kaplinghat}},\ and\
  \bibinfo {author} {\bibfnamefont {L.}~\bibnamefont {Sagunski}},\ }\bibfield
  {title} {\bibinfo {title} {{Universal gravothermal evolution of isolated
  self-interacting dark matter halos for velocity-dependent cross-sections}},\
  }\href {https://doi.org/10.1093/mnras/stad1705} {\bibfield  {journal}
  {\bibinfo  {journal} {Mon. Not. Roy. Astron. Soc.}\ }\textbf {\bibinfo
  {volume} {523}},\ \bibinfo {pages} {4786} (\bibinfo {year} {2023})},\ \Eprint
  {https://arxiv.org/abs/2204.06568} {arXiv:2204.06568 [astro-ph.GA]}
  \BibitemShut {NoStop}%
\bibitem [{\citenamefont {Yang}\ \emph
  {et~al.}(2023{\natexlab{b}})\citenamefont {Yang}, \citenamefont {Du},
  \citenamefont {Zeng}, \citenamefont {Benson}, \citenamefont {Jiang},
  \citenamefont {Nadler},\ and\ \citenamefont {Peter}}]{Yang:2022zkd}%
  \BibitemOpen
  \bibfield  {author} {\bibinfo {author} {\bibfnamefont {S.}~\bibnamefont
  {Yang}}, \bibinfo {author} {\bibfnamefont {X.}~\bibnamefont {Du}}, \bibinfo
  {author} {\bibfnamefont {Z.~C.}\ \bibnamefont {Zeng}}, \bibinfo {author}
  {\bibfnamefont {A.}~\bibnamefont {Benson}}, \bibinfo {author} {\bibfnamefont
  {F.}~\bibnamefont {Jiang}}, \bibinfo {author} {\bibfnamefont {E.~O.}\
  \bibnamefont {Nadler}},\ and\ \bibinfo {author} {\bibfnamefont {A.~H.~G.}\
  \bibnamefont {Peter}},\ }\bibfield  {title} {\bibinfo {title} {{Gravothermal
  Solutions of SIDM Halos: Mapping from Constant to Velocity-dependent Cross
  Section}},\ }\href {https://doi.org/10.3847/1538-4357/acbd49} {\bibfield
  {journal} {\bibinfo  {journal} {Astrophys. J.}\ }\textbf {\bibinfo {volume}
  {946}},\ \bibinfo {pages} {47} (\bibinfo {year} {2023}{\natexlab{b}})},\
  \Eprint {https://arxiv.org/abs/2205.02957} {arXiv:2205.02957 [astro-ph.CO]}
  \BibitemShut {NoStop}%
\bibitem [{\citenamefont {Gad-Nasr}\ \emph {et~al.}(2024)\citenamefont
  {Gad-Nasr}, \citenamefont {Boddy}, \citenamefont {Kaplinghat}, \citenamefont
  {Outmezguine},\ and\ \citenamefont {Sagunski}}]{Gad-Nasr:2023gvf}%
  \BibitemOpen
  \bibfield  {author} {\bibinfo {author} {\bibfnamefont {S.}~\bibnamefont
  {Gad-Nasr}}, \bibinfo {author} {\bibfnamefont {K.~K.}\ \bibnamefont {Boddy}},
  \bibinfo {author} {\bibfnamefont {M.}~\bibnamefont {Kaplinghat}}, \bibinfo
  {author} {\bibfnamefont {N.~J.}\ \bibnamefont {Outmezguine}},\ and\ \bibinfo
  {author} {\bibfnamefont {L.}~\bibnamefont {Sagunski}},\ }\bibfield  {title}
  {\bibinfo {title} {{On the late-time evolution of velocity-dependent
  self-interacting dark matter halos}},\ }\href
  {https://doi.org/10.1088/1475-7516/2024/05/131} {\bibfield  {journal}
  {\bibinfo  {journal} {JCAP}\ }\textbf {\bibinfo {volume} {05}},\ \bibinfo
  {pages} {131}},\ \Eprint {https://arxiv.org/abs/2312.09296} {arXiv:2312.09296
  [astro-ph.GA]} \BibitemShut {NoStop}%
\bibitem [{\citenamefont {Dave}\ \emph {et~al.}(2001)\citenamefont {Dave},
  \citenamefont {Spergel}, \citenamefont {Steinhardt},\ and\ \citenamefont
  {Wandelt}}]{Dave:2000ar}%
  \BibitemOpen
  \bibfield  {author} {\bibinfo {author} {\bibfnamefont {R.}~\bibnamefont
  {Dave}}, \bibinfo {author} {\bibfnamefont {D.~N.}\ \bibnamefont {Spergel}},
  \bibinfo {author} {\bibfnamefont {P.~J.}\ \bibnamefont {Steinhardt}},\ and\
  \bibinfo {author} {\bibfnamefont {B.~D.}\ \bibnamefont {Wandelt}},\
  }\bibfield  {title} {\bibinfo {title} {{Halo properties in cosmological
  simulations of selfinteracting cold dark matter}},\ }\href
  {https://doi.org/10.1086/318417} {\bibfield  {journal} {\bibinfo  {journal}
  {Astrophys. J.}\ }\textbf {\bibinfo {volume} {547}},\ \bibinfo {pages} {574}
  (\bibinfo {year} {2001})},\ \Eprint {https://arxiv.org/abs/astro-ph/0006218}
  {arXiv:astro-ph/0006218} \BibitemShut {NoStop}%
\bibitem [{\citenamefont {Kummer}\ \emph {et~al.}(2019)\citenamefont {Kummer},
  \citenamefont {Br\"uggen}, \citenamefont {Dolag}, \citenamefont
  {Kahlhoefer},\ and\ \citenamefont {Schmidt-Hoberg}}]{Kummer:2019yrb}%
  \BibitemOpen
  \bibfield  {author} {\bibinfo {author} {\bibfnamefont {J.}~\bibnamefont
  {Kummer}}, \bibinfo {author} {\bibfnamefont {M.}~\bibnamefont {Br\"uggen}},
  \bibinfo {author} {\bibfnamefont {K.}~\bibnamefont {Dolag}}, \bibinfo
  {author} {\bibfnamefont {F.}~\bibnamefont {Kahlhoefer}},\ and\ \bibinfo
  {author} {\bibfnamefont {K.}~\bibnamefont {Schmidt-Hoberg}},\ }\bibfield
  {title} {\bibinfo {title} {{Simulations of core formation for frequent dark
  matter self-interactions}},\ }\href {https://doi.org/10.1093/mnras/stz1261}
  {\bibfield  {journal} {\bibinfo  {journal} {Mon. Not. Roy. Astron. Soc.}\
  }\textbf {\bibinfo {volume} {487}},\ \bibinfo {pages} {354} (\bibinfo {year}
  {2019})},\ \Eprint {https://arxiv.org/abs/1902.02330} {arXiv:1902.02330
  [astro-ph.CO]} \BibitemShut {NoStop}%
\bibitem [{\citenamefont {Fischer}\ \emph {et~al.}(2021)\citenamefont
  {Fischer}, \citenamefont {Br\"uggen}, \citenamefont {Schmidt-Hoberg},
  \citenamefont {Dolag}, \citenamefont {Kahlhoefer}, \citenamefont {Ragagnin},\
  and\ \citenamefont {Robertson}}]{Fischer:2020uxh}%
  \BibitemOpen
  \bibfield  {author} {\bibinfo {author} {\bibfnamefont {M.~S.}\ \bibnamefont
  {Fischer}}, \bibinfo {author} {\bibfnamefont {M.}~\bibnamefont {Br\"uggen}},
  \bibinfo {author} {\bibfnamefont {K.}~\bibnamefont {Schmidt-Hoberg}},
  \bibinfo {author} {\bibfnamefont {K.}~\bibnamefont {Dolag}}, \bibinfo
  {author} {\bibfnamefont {F.}~\bibnamefont {Kahlhoefer}}, \bibinfo {author}
  {\bibfnamefont {A.}~\bibnamefont {Ragagnin}},\ and\ \bibinfo {author}
  {\bibfnamefont {A.}~\bibnamefont {Robertson}},\ }\bibfield  {title} {\bibinfo
  {title} {{N-body simulations of dark matter with frequent
  self-interactions}},\ }\href {https://doi.org/10.1093/mnras/stab1198}
  {\bibfield  {journal} {\bibinfo  {journal} {Mon. Not. Roy. Astron. Soc.}\
  }\textbf {\bibinfo {volume} {505}},\ \bibinfo {pages} {851} (\bibinfo {year}
  {2021})},\ \Eprint {https://arxiv.org/abs/2012.10277} {arXiv:2012.10277
  [astro-ph.CO]} \BibitemShut {NoStop}%
\bibitem [{\citenamefont {Mace}\ \emph {et~al.}(2024)\citenamefont {Mace},
  \citenamefont {Zeng}, \citenamefont {Peter}, \citenamefont {Du},
  \citenamefont {Yang}, \citenamefont {Benson},\ and\ \citenamefont
  {Vogelsberger}}]{Mace:2024uze}%
  \BibitemOpen
  \bibfield  {author} {\bibinfo {author} {\bibfnamefont {C.}~\bibnamefont
  {Mace}}, \bibinfo {author} {\bibfnamefont {Z.~C.}\ \bibnamefont {Zeng}},
  \bibinfo {author} {\bibfnamefont {A.~H.~G.}\ \bibnamefont {Peter}}, \bibinfo
  {author} {\bibfnamefont {X.}~\bibnamefont {Du}}, \bibinfo {author}
  {\bibfnamefont {S.}~\bibnamefont {Yang}}, \bibinfo {author} {\bibfnamefont
  {A.}~\bibnamefont {Benson}},\ and\ \bibinfo {author} {\bibfnamefont
  {M.}~\bibnamefont {Vogelsberger}},\ }\bibfield  {title} {\bibinfo {title}
  {{Convergence tests of self-interacting dark matter simulations}},\ }\href
  {https://doi.org/10.1103/PhysRevD.110.123024} {\bibfield  {journal} {\bibinfo
   {journal} {Phys. Rev. D}\ }\textbf {\bibinfo {volume} {110}},\ \bibinfo
  {pages} {123024} (\bibinfo {year} {2024})},\ \Eprint
  {https://arxiv.org/abs/2402.01604} {arXiv:2402.01604 [astro-ph.GA]}
  \BibitemShut {NoStop}%
\bibitem [{\citenamefont {Palubski}\ \emph {et~al.}(2024)\citenamefont
  {Palubski}, \citenamefont {Slone}, \citenamefont {Kaplinghat}, \citenamefont
  {Lisanti},\ and\ \citenamefont {Jiang}}]{Palubski:2024ibb}%
  \BibitemOpen
  \bibfield  {author} {\bibinfo {author} {\bibfnamefont {I.}~\bibnamefont
  {Palubski}}, \bibinfo {author} {\bibfnamefont {O.}~\bibnamefont {Slone}},
  \bibinfo {author} {\bibfnamefont {M.}~\bibnamefont {Kaplinghat}}, \bibinfo
  {author} {\bibfnamefont {M.}~\bibnamefont {Lisanti}},\ and\ \bibinfo {author}
  {\bibfnamefont {F.}~\bibnamefont {Jiang}},\ }\bibfield  {title} {\bibinfo
  {title} {{Numerical challenges in modeling gravothermal collapse in
  Self-Interacting Dark Matter halos}},\ }\href
  {https://doi.org/10.1088/1475-7516/2024/09/074} {\bibfield  {journal}
  {\bibinfo  {journal} {JCAP}\ }\textbf {\bibinfo {volume} {09}},\ \bibinfo
  {pages} {074}},\ \Eprint {https://arxiv.org/abs/2402.12452} {arXiv:2402.12452
  [astro-ph.CO]} \BibitemShut {NoStop}%
\bibitem [{\citenamefont {Fischer}\ \emph {et~al.}(2024)\citenamefont
  {Fischer}, \citenamefont {Dolag},\ and\ \citenamefont
  {Yu}}]{Fischer:2024eaz}%
  \BibitemOpen
  \bibfield  {author} {\bibinfo {author} {\bibfnamefont {M.~S.}\ \bibnamefont
  {Fischer}}, \bibinfo {author} {\bibfnamefont {K.}~\bibnamefont {Dolag}},\
  and\ \bibinfo {author} {\bibfnamefont {H.-B.}\ \bibnamefont {Yu}},\
  }\bibfield  {title} {\bibinfo {title} {{Numerical challenges for energy
  conservation in N-body simulations of collapsing self-interacting dark matter
  halos}},\ }\href {https://doi.org/10.1051/0004-6361/202449849} {\bibfield
  {journal} {\bibinfo  {journal} {Astron. Astrophys.}\ }\textbf {\bibinfo
  {volume} {689}},\ \bibinfo {pages} {A300} (\bibinfo {year} {2024})},\ \Eprint
  {https://arxiv.org/abs/2403.00739} {arXiv:2403.00739 [astro-ph.CO]}
  \BibitemShut {NoStop}%
\bibitem [{\citenamefont {Mace}\ \emph {et~al.}(2025)\citenamefont {Mace},
  \citenamefont {Yang}, \citenamefont {Zeng}, \citenamefont {Peter},
  \citenamefont {Du},\ and\ \citenamefont {Benson}}]{Mace:2025fuz}%
  \BibitemOpen
  \bibfield  {author} {\bibinfo {author} {\bibfnamefont {C.}~\bibnamefont
  {Mace}}, \bibinfo {author} {\bibfnamefont {S.}~\bibnamefont {Yang}}, \bibinfo
  {author} {\bibfnamefont {Z.~C.}\ \bibnamefont {Zeng}}, \bibinfo {author}
  {\bibfnamefont {A.~H.~G.}\ \bibnamefont {Peter}}, \bibinfo {author}
  {\bibfnamefont {X.}~\bibnamefont {Du}},\ and\ \bibinfo {author}
  {\bibfnamefont {A.}~\bibnamefont {Benson}},\ }\bibfield  {title} {\bibinfo
  {title} {{Calibrating the SIDM Gravothermal Catastrophe with N-body
  Simulations}},\ }\href@noop {} {\  (\bibinfo {year} {2025})},\ \Eprint
  {https://arxiv.org/abs/2504.13004} {arXiv:2504.13004 [astro-ph.GA]}
  \BibitemShut {NoStop}%
\bibitem [{\citenamefont {Fischer}\ \emph {et~al.}(2025)\citenamefont
  {Fischer}, \citenamefont {Yu},\ and\ \citenamefont
  {Dolag}}]{Fischer:2025rky}%
  \BibitemOpen
  \bibfield  {author} {\bibinfo {author} {\bibfnamefont {M.~S.}\ \bibnamefont
  {Fischer}}, \bibinfo {author} {\bibfnamefont {H.-B.}\ \bibnamefont {Yu}},\
  and\ \bibinfo {author} {\bibfnamefont {K.}~\bibnamefont {Dolag}},\ }\bibfield
   {title} {\bibinfo {title} {{Accurately simulating core-collapse
  self-interacting dark matter halos}},\ }\href@noop {} {\  (\bibinfo {year}
  {2025})},\ \Eprint {https://arxiv.org/abs/2506.06269} {arXiv:2506.06269
  [astro-ph.CO]} \BibitemShut {NoStop}%
\bibitem [{\citenamefont {{Eddington}}(1916)}]{Eddington:1916}%
  \BibitemOpen
  \bibfield  {author} {\bibinfo {author} {\bibfnamefont {A.~S.}\ \bibnamefont
  {{Eddington}}},\ }\bibfield  {title} {\bibinfo {title} {{The distribution of
  stars in globular clusters}},\ }\href
  {https://doi.org/10.1093/mnras/76.7.572} {\bibfield  {journal} {\bibinfo
  {journal} {Mon. Not. Roy. Astron. Soc.}\ }\textbf {\bibinfo {volume} {76}},\
  \bibinfo {pages} {572} (\bibinfo {year} {1916})}\BibitemShut {NoStop}%
\bibitem [{\citenamefont {Binney}\ and\ \citenamefont
  {Tremaine}(2008)}]{Binney:2008}%
  \BibitemOpen
  \bibfield  {author} {\bibinfo {author} {\bibfnamefont {J.}~\bibnamefont
  {Binney}}\ and\ \bibinfo {author} {\bibfnamefont {S.}~\bibnamefont
  {Tremaine}},\ }\href@noop {} {\emph {\bibinfo {title} {Galactic Dynamics}}},\
  \bibinfo {edition} {2nd}\ ed.\ (\bibinfo  {publisher} {Princeton University
  Press},\ \bibinfo {address} {Princeton, NJ},\ \bibinfo {year}
  {2008})\BibitemShut {NoStop}%
\bibitem [{\citenamefont {Wojtak}\ \emph {et~al.}(2013)\citenamefont {Wojtak},
  \citenamefont {Gottloeber},\ and\ \citenamefont {Klypin}}]{Wojtak:2013eia}%
  \BibitemOpen
  \bibfield  {author} {\bibinfo {author} {\bibfnamefont {R.}~\bibnamefont
  {Wojtak}}, \bibinfo {author} {\bibfnamefont {S.}~\bibnamefont {Gottloeber}},\
  and\ \bibinfo {author} {\bibfnamefont {A.}~\bibnamefont {Klypin}},\
  }\bibfield  {title} {\bibinfo {title} {{Orbital anisotropy in cosmological
  haloes revisited}},\ }\href {https://doi.org/10.1093/mnras/stt1113}
  {\bibfield  {journal} {\bibinfo  {journal} {Mon. Not. Roy. Astron. Soc.}\
  }\textbf {\bibinfo {volume} {434}},\ \bibinfo {pages} {1576} (\bibinfo {year}
  {2013})},\ \Eprint {https://arxiv.org/abs/1303.2056} {arXiv:1303.2056
  [astro-ph.CO]} \BibitemShut {NoStop}%
\bibitem [{\citenamefont {He}\ \emph {et~al.}(2024)\citenamefont {He} \emph
  {et~al.}}]{He:2024gvw}%
  \BibitemOpen
  \bibfield  {author} {\bibinfo {author} {\bibfnamefont {J.}~\bibnamefont {He}}
  \emph {et~al.},\ }\bibfield  {title} {\bibinfo {title} {{How Do the Velocity
  Anisotropies of Halo Stars, Dark Matter, and Satellite Galaxies Depend on
  Host Halo Properties?}},\ }\href {https://doi.org/10.3847/1538-4357/ad8882}
  {\bibfield  {journal} {\bibinfo  {journal} {Astrophys. J.}\ }\textbf
  {\bibinfo {volume} {976}},\ \bibinfo {pages} {187} (\bibinfo {year}
  {2024})},\ \Eprint {https://arxiv.org/abs/2407.14827} {arXiv:2407.14827
  [astro-ph.GA]} \BibitemShut {NoStop}%
\bibitem [{\citenamefont {Lemze}\ \emph {et~al.}(2012)\citenamefont {Lemze},
  \citenamefont {Wagner}, \citenamefont {Rephaeli}, \citenamefont {Sadeh},
  \citenamefont {Norman}, \citenamefont {Barkana}, \citenamefont {Broadhurst},
  \citenamefont {Ford},\ and\ \citenamefont {Postman}}]{Lemze_2012}%
  \BibitemOpen
  \bibfield  {author} {\bibinfo {author} {\bibfnamefont {D.}~\bibnamefont
  {Lemze}}, \bibinfo {author} {\bibfnamefont {R.}~\bibnamefont {Wagner}},
  \bibinfo {author} {\bibfnamefont {Y.}~\bibnamefont {Rephaeli}}, \bibinfo
  {author} {\bibfnamefont {S.}~\bibnamefont {Sadeh}}, \bibinfo {author}
  {\bibfnamefont {M.~L.}\ \bibnamefont {Norman}}, \bibinfo {author}
  {\bibfnamefont {R.}~\bibnamefont {Barkana}}, \bibinfo {author} {\bibfnamefont
  {T.}~\bibnamefont {Broadhurst}}, \bibinfo {author} {\bibfnamefont
  {H.}~\bibnamefont {Ford}},\ and\ \bibinfo {author} {\bibfnamefont
  {M.}~\bibnamefont {Postman}},\ }\bibfield  {title} {\bibinfo {title}
  {Profiles of dark matter velocity anisotropy in simulated clusters},\ }\href
  {https://doi.org/10.1088/0004-637x/752/2/141} {\bibfield  {journal} {\bibinfo
   {journal} {The Astrophysical Journal}\ }\textbf {\bibinfo {volume} {752}},\
  \bibinfo {pages} {141} (\bibinfo {year} {2012})}\BibitemShut {NoStop}%
\bibitem [{\citenamefont {Biviano}\ \emph {et~al.}(2013)\citenamefont {Biviano}
  \emph {et~al.}}]{Biviano:2013eia}%
  \BibitemOpen
  \bibfield  {author} {\bibinfo {author} {\bibfnamefont {A.}~\bibnamefont
  {Biviano}} \emph {et~al.},\ }\bibfield  {title} {\bibinfo {title}
  {{CLASH-VLT: The mass, velocity-anisotropy, and pseudo-phase-space density
  profiles of the z=0.44 galaxy cluster MACS 1206.2-0847}},\ }\href
  {https://doi.org/10.1051/0004-6361/201321955} {\bibfield  {journal} {\bibinfo
   {journal} {Astron. Astrophys.}\ }\textbf {\bibinfo {volume} {558}},\
  \bibinfo {pages} {A1} (\bibinfo {year} {2013})},\ \Eprint
  {https://arxiv.org/abs/1307.5867} {arXiv:1307.5867 [astro-ph.CO]}
  \BibitemShut {NoStop}%
\bibitem [{\citenamefont {Kamionkowski}\ and\ \citenamefont
  {Sigurdson}(2025)}]{Kamionkowski:2025uae}%
  \BibitemOpen
  \bibfield  {author} {\bibinfo {author} {\bibfnamefont {M.}~\bibnamefont
  {Kamionkowski}}\ and\ \bibinfo {author} {\bibfnamefont {K.}~\bibnamefont
  {Sigurdson}},\ }\bibfield  {title} {\bibinfo {title} {{Evolution of
  self-gravitating spherical dark-matter halos with and without new physics}},\
  }\href@noop {} {\  (\bibinfo {year} {2025})},\ \Eprint
  {https://arxiv.org/abs/2504.13996} {arXiv:2504.13996 [astro-ph.GA]}
  \BibitemShut {NoStop}%
\bibitem [{\citenamefont {Kamionkowski}\ \emph {et~al.}(2025)\citenamefont
  {Kamionkowski}, \citenamefont {Sigurdson},\ and\ \citenamefont
  {Slone}}]{Kamionkowski:2025fat}%
  \BibitemOpen
  \bibfield  {author} {\bibinfo {author} {\bibfnamefont {M.}~\bibnamefont
  {Kamionkowski}}, \bibinfo {author} {\bibfnamefont {K.}~\bibnamefont
  {Sigurdson}},\ and\ \bibinfo {author} {\bibfnamefont {O.}~\bibnamefont
  {Slone}},\ }\bibfield  {title} {\bibinfo {title} {{Numerical evolution of
  self-gravitating halos of self-interacting dark matter}},\ }\href@noop {} {\
  (\bibinfo {year} {2025})},\ \Eprint {https://arxiv.org/abs/2506.04334}
  {arXiv:2506.04334 [astro-ph.CO]} \BibitemShut {NoStop}%
\bibitem [{\citenamefont {Gurian}\ and\ \citenamefont
  {May}(2025)}]{Gurian:2025zpc}%
  \BibitemOpen
  \bibfield  {author} {\bibinfo {author} {\bibfnamefont {J.}~\bibnamefont
  {Gurian}}\ and\ \bibinfo {author} {\bibfnamefont {S.}~\bibnamefont {May}},\
  }\bibfield  {title} {\bibinfo {title} {{Core Collapse Beyond the Fluid
  Approximation: The Late Evolution of Self-Interacting Dark Matter Halos}},\
  }\href@noop {} {\  (\bibinfo {year} {2025})},\ \Eprint
  {https://arxiv.org/abs/2505.15903} {arXiv:2505.15903 [astro-ph.CO]}
  \BibitemShut {NoStop}%
\bibitem [{\citenamefont {Jiang}\ and\ \citenamefont
  {Ossipkov}(2007)}]{Jiang:2007vsd}%
  \BibitemOpen
  \bibfield  {author} {\bibinfo {author} {\bibfnamefont {Z.}~\bibnamefont
  {Jiang}}\ and\ \bibinfo {author} {\bibfnamefont {L.}~\bibnamefont
  {Ossipkov}},\ }\bibfield  {title} {\bibinfo {title} {{Anisotropic
  distribution functions for spherical galaxies}},\ }\href
  {https://doi.org/10.1007/s10569-006-9062-5} {\bibfield  {journal} {\bibinfo
  {journal} {Celest. Mech. Dyn. Astron.}\ }\textbf {\bibinfo {volume} {97}},\
  \bibinfo {pages} {249} (\bibinfo {year} {2007})},\ \Eprint
  {https://arxiv.org/abs/0812.4898} {arXiv:0812.4898 [astro-ph]} \BibitemShut
  {NoStop}%
\bibitem [{\citenamefont {Ossipkov}(1979)}]{Ossipkov1979}%
  \BibitemOpen
  \bibfield  {author} {\bibinfo {author} {\bibfnamefont {L.~P.}\ \bibnamefont
  {Ossipkov}},\ }\bibfield  {title} {\bibinfo {title} {Some problems of the
  theory of self-consistent models for star clusters},\ }in\ \href@noop {}
  {\emph {\bibinfo {booktitle} {Star Clusters}}}\ (\bibinfo  {publisher} {Urals
  University Press},\ \bibinfo {address} {Sverdlovsk},\ \bibinfo {year}
  {1979})\ pp.\ \bibinfo {pages} {72--89}\BibitemShut {NoStop}%
\bibitem [{\citenamefont {Merritt}(1985)}]{Merritt1985}%
  \BibitemOpen
  \bibfield  {author} {\bibinfo {author} {\bibfnamefont {D.}~\bibnamefont
  {Merritt}},\ }\bibfield  {title} {\bibinfo {title} {Spherical stellar systems
  with spheroidal velocity distributions},\ }\href
  {https://doi.org/10.1086/113810} {\bibfield  {journal} {\bibinfo  {journal}
  {Astronomical Journal}\ }\textbf {\bibinfo {volume} {90}},\ \bibinfo {pages}
  {1027} (\bibinfo {year} {1985})}\BibitemShut {NoStop}%
\bibitem [{\citenamefont {Dejonghe}(1987)}]{Dejonghe1987}%
  \BibitemOpen
  \bibfield  {author} {\bibinfo {author} {\bibfnamefont {H.}~\bibnamefont
  {Dejonghe}},\ }\bibfield  {title} {\bibinfo {title} {A completely analytical
  family of anisotropic {Plummer} models},\ }\href
  {https://doi.org/10.1093/mnras/224.1.13} {\bibfield  {journal} {\bibinfo
  {journal} {Monthly Notices of the Royal Astronomical Society}\ }\textbf
  {\bibinfo {volume} {224}},\ \bibinfo {pages} {13} (\bibinfo {year}
  {1987})}\BibitemShut {NoStop}%
\bibitem [{\citenamefont {Baes}\ and\ \citenamefont
  {Dejonghe}(2002)}]{Baes:2002tw}%
  \BibitemOpen
  \bibfield  {author} {\bibinfo {author} {\bibfnamefont {M.}~\bibnamefont
  {Baes}}\ and\ \bibinfo {author} {\bibfnamefont {H.}~\bibnamefont
  {Dejonghe}},\ }\bibfield  {title} {\bibinfo {title} {{The Hernquist model
  revisited: Completely analytical anisotropic dynamical models}},\ }\href
  {https://doi.org/10.1051/0004-6361:20021064} {\bibfield  {journal} {\bibinfo
  {journal} {Astron. Astrophys.}\ }\textbf {\bibinfo {volume} {393}},\ \bibinfo
  {pages} {485} (\bibinfo {year} {2002})},\ \Eprint
  {https://arxiv.org/abs/astro-ph/0207233} {arXiv:astro-ph/0207233}
  \BibitemShut {NoStop}%
\bibitem [{\citenamefont {{An}}\ and\ \citenamefont {{Evans}}(2006)}]{AnEvans}%
  \BibitemOpen
  \bibfield  {author} {\bibinfo {author} {\bibfnamefont {J.~H.}\ \bibnamefont
  {{An}}}\ and\ \bibinfo {author} {\bibfnamefont {N.~W.}\ \bibnamefont
  {{Evans}}},\ }\bibfield  {title} {\bibinfo {title} {{Galaxy Models with
  Tangentially Anisotropic Velocity Distributions}},\ }\href
  {https://doi.org/10.1086/499305} {\bibfield  {journal} {\bibinfo  {journal}
  {Astron.\ J.}\ }\textbf {\bibinfo {volume} {131}},\ \bibinfo {pages} {782}
  (\bibinfo {year} {2006})},\ \Eprint {https://arxiv.org/abs/astro-ph/0501092}
  {arXiv:astro-ph/0501092 [astro-ph]} \BibitemShut {NoStop}%
\bibitem [{\citenamefont {Hernquist}(1990)}]{Hernquist:1990be}%
  \BibitemOpen
  \bibfield  {author} {\bibinfo {author} {\bibfnamefont {L.}~\bibnamefont
  {Hernquist}},\ }\bibfield  {title} {\bibinfo {title} {{An Analytical Model
  for Spherical Galaxies and Bulges}},\ }\href {https://doi.org/10.1086/168845}
  {\bibfield  {journal} {\bibinfo  {journal} {Astrophys. J.}\ }\textbf
  {\bibinfo {volume} {356}},\ \bibinfo {pages} {359} (\bibinfo {year}
  {1990})}\BibitemShut {NoStop}%
\bibitem [{\citenamefont {Navarro}\ \emph {et~al.}(1997)\citenamefont
  {Navarro}, \citenamefont {Frenk},\ and\ \citenamefont
  {White}}]{Navarro:1996gj}%
  \BibitemOpen
  \bibfield  {author} {\bibinfo {author} {\bibfnamefont {J.~F.}\ \bibnamefont
  {Navarro}}, \bibinfo {author} {\bibfnamefont {C.~S.}\ \bibnamefont {Frenk}},\
  and\ \bibinfo {author} {\bibfnamefont {S.~D.~M.}\ \bibnamefont {White}},\
  }\bibfield  {title} {\bibinfo {title} {{A Universal density profile from
  hierarchical clustering}},\ }\href {https://doi.org/10.1086/304888}
  {\bibfield  {journal} {\bibinfo  {journal} {Astrophys. J.}\ }\textbf
  {\bibinfo {volume} {490}},\ \bibinfo {pages} {493} (\bibinfo {year}
  {1997})},\ \Eprint {https://arxiv.org/abs/astro-ph/9611107}
  {arXiv:astro-ph/9611107} \BibitemShut {NoStop}%
\end{thebibliography}%

\end{document}